# Ultimate Performance of Thin-Film Lithium Niobate Mach-Zehnder Modulators

S. Behzadfar, F. Karami, P. Kulkarni, and S. Fathpour

***Abstract*—In thin-film lithium niobate Mach-Zehnder modulators, half-wave voltage and bandwidth are tied to the same design parameters; hence, improving one usually costs the other. Several methods have been pursued in the past to alleviate this tradeoff. Specifically, the bias voltage can be reduced through non-centered placement of optical waveguides with respect to electrodes, slow-wave optical gratings, and high-permittivity cladding layers. Bandwidth can be enhanced through implementing periodic segmented electrodes and exploiting low-permittivity bottom cladding and substrate materials. The objective of this work is to study the augmentative impact of these methods. We pursue a rigorous modeling and optimization approach to achieve this objective, estimate the impact of each augmentation, and evaluate the ultimate performance of TFLN MZMs based on these methods. It is shown that a co-designed structure that incorporates all these enhancing methods can attain a drive voltage of 0.84 V and a modulation bandwidth of almost 220 GHz.**

## I. Introduction

High-speed photonic links for data centers, telecommunications, microwave-photonic signal processors, optical interconnects, and 5G and 6G front-haul systems call for electro-optic modulators (EOMs) that pair sub-volt drive voltages with modulation bandwidths reaching into the millimeter-wave and sub-terahertz range [1], [2]. EOMs increasingly carry system-level functions as well, among them the cancellation of fiber dispersion and laser noise through differential phase-diversity schemes [3]. Compact EOMs based on Thin-Film Lithium Niobate (TFLN) address many of these demands simultaneously [2], [4]. Ion slicing and wafer bonding were the turning point in the development of TFLN integrated photonics technology [5]. By shrinking the optical mode areas by roughly thirtyfold relative to titanium-diffused or proton-exchanged lithium niobate (LN) waveguides, electrode gaps of a few micrometers became reachable. In addition, the $V_\pi \cdot L$ decreases markedly. The development of TFLN wafers led to the first demonstrations of high-speed Mach-Zehnder modulators (MZMs) [6] and paved the path to the 100 GHz electro-optic (EO) bandwidth milestone [7]. Despite this rapid progress, CMOS-compatible TFLN MZMs, i.e., half-wave voltage, $V_\pi$, of < 1 V, and EO bandwidths of > 100 GHz, have been more challenging to attain simultaneously, which is the subject of this work.

In this work, we first review the basic formulations to model the drive voltage and modulation bandwidth of TFLN EOMs. Then, five major methods for higher performance are described. These techniques include a displaced waveguide, a high-permittivity ($\varepsilon$) cladding material, capacitively loaded traveling-wave electrodes, low-$\varepsilon$ substrates, and slow-wave group-index enhancement. For each case, we summarize the benefits and costs and the reported performances in the literature. We then study a co-designed EOM structure, where all these routes are augmented to contemplate what the ultimate performance limit of TFLN MZMs could be, under other practical considerations, such as negligible optical loss induced by metallic electrodes. We also study and compare the intermediate cases, where each enhancing technique is added to the most basic conventional TFLN MZMs. It is ultimately shown that, fabrication challenges notwithstanding, an air-undercut TFLN displaced slow-light waveguide on a quartz substrate, with a glycerol upper cladding and an optimized periodic segmented electrode, can attain a half-wave voltage of 0.84 V and an EO modulation bandwidth of $\sim 219$ GHz.

## II. Basic Relations For $V_\pi$ And EO Transfer Function

This section models MZMs with respect to three major design goals. The first goal is to minimize $V_\pi$, which is set by the electrode gap, the optical group index, and the EO overlap integral. The second goal is to achieve a broadband EO bandwidth, which depends on the radio-frequency (RF) loss of the electrodes, the microwave-optical velocity mismatch, and the line impedance. Third, the optical loss must be kept as low as possible; within the design space, this loss is governed by metal absorption, which rises steeply at narrow electrode gaps. The central difficulty of EOM design is that these goals are not independent; typically, any approach that improves one may disturb the others.

This work was supported by the National Science Foundation Industry–University Cooperative Research Centers (IUCRC) EPICA Program. *(Corresponding author: Sasan Fathpour.)*

Shiva Behzadfar, Fatemeh Karami, and Pooja Kulkarni are with CREOL, The College of Optics and Photonics, University of Central Florida, Orlando, FL 32816 USA (e-mail: shiva.behzadfar@ucf.edu; pooja.kulkarni@ucf.edu; fatemeh.karami@ucf.edu).

Sasan Fathpour is with CREOL, The College of Optics and Photonics, University of Central Florida, Orlando, FL 32816 USA, and also with the Department of Electrical and Computer Engineering, University of Central Florida, Orlando, FL 32816 USA (e-mail: fathpour@creol.ucf.edu).

$V_\pi$ occurs when the EO phase shift accumulated over the interaction length $L$, equals π. With an electrode gap, $G$ (see Fig. 1(a)), optical group index, $n_g$, and EO overlap integral, Γ, the half-wave-voltage-length product, $V_\pi.L$, for the push-pull configuration and X-cut wafers, is [8]

$$V_\pi \cdot L = \frac{\lambda\, n_{eff}^2\, G}{2\, n_e^4\, n_g r_{33}\, \Gamma}, \tag{1}$$

where λ is the free-space wavelength, $n_{\text{eff}}$ the optical mode effective index, $n_{\text{e}}$ the extraordinary index of bulk LN, and $r_{33}$ the Pockels coefficient. The EO overlap is the cross-sectional integral of the RF field with the optical mode, $E_{\text{opt}}(x,y)$, inside LN, and is given by

$$\Gamma = \frac{\iint_{\text{LN}} f_z(x,y)\, |E_{\text{opt}}(x,y)|^2\, \mathrm{d}x\, \mathrm{d}y}{\iint_{\infty} |E_{\text{opt}}(x,y)|^2\, \mathrm{d}x\, \mathrm{d}y}, \tag{2}$$

where $f_z(x,y)$ is a normalized and unitless modulating RF field along the lateral direction (corresponding to $r_{33}$).

The frequency-dependent EO response of a traveling-wave electrode EOM, with $\omega_{\text{RF}} = 2\pi f$, is given by the transfer function [9]

$$H(\omega_{RF}) = \frac{Z_{\text{in}}}{Z_{\text{in}} + Z_S} \frac{1}{\mathrm{e}^{\gamma L} + \Gamma_{\text{L}}\, \mathrm{e}^{-\gamma L}} \tag{3}$$

$$\times \left[ \frac{\sinh(AL/2)\ \mathrm{e}^{-AL/2}}{A/2} + \Gamma_L \frac{\sinh(BL/2)\ \mathrm{e}^{-BL/2}}{B/2} \right]$$

$$A = -\alpha_{\text{RF}} - \mathrm{j}\omega_{\text{RF}} \frac{(n_{\text{RF}} - n_{\text{eff}})}{c},$$

$$B = +\alpha_{\text{RF}} + \mathrm{j}\omega_{\text{RF}} \frac{(n_{\text{RF}} + n_{\text{eff}})}{c}$$

Here, $Z_{\text{in}}$ is the input impedance seen at the input of the transmission line, $Z_S$ the source impedance, and $\Gamma_{\text{L}} = (Z_L - Z_0)/(Z_L + Z_0)$ is the reflection coefficient at the terminating load with impedance $Z_L$, while $Z_0$ is the transmission line's characteristic impedance. Furthermore, $\gamma = \alpha_{\text{RF}} + \mathrm{j}\beta_{\text{RF}}$ is the complex RF propagation constant, $n_{RF}$ is the RF effective refractive index, and $c$ is the speed of the electromagnetic wave in a vacuum. The velocity mismatch between the microwave and optical waves is set by the difference between the RF and optical effective indices, denoted by $\Delta n = (n_{\text{RF}} - n_{\text{eff}})$.

The 3-dB EO modulation bandwidth is the frequency at which $|H(\omega_{\text{RF}})|$ falls to one-half of its DC value, and its two main limiting cases are set by either the RF loss or velocity mismatch in impedance-matched circuits. RF attenuation is the sum of conductor (ohmic), dielectric, and radiation contributions, which can increase with frequency. For example, in the case of coplanar waveguides (CPW), the dominant conductor term carries the skin effect and is proportional to $\sqrt{f}$ [9]. The velocity-mismatch-limited bandwidth is

$$f_{\text{3dB,eo}} \approx \frac{0.60\, c}{|\Delta n(f)| L}, \tag{4}$$

where the 0.60 factor is the argument of the sinc$^2$ function at the half-power point. Similarly, the loss-limited 3-dB bandwidth occurs when $\alpha_{\text{RF}} L \approx 3.2$, and under the Wheeler approximation, i.e., when the conductor thickness of a CPW is much greater than its skin depth, is given by

$$f_{\text{3dB,eo}} \approx \left(\frac{3.2}{\alpha_{c0} L}\right)^2, \tag{5}$$

where $\alpha_{c0}$ is the conductor-loss coefficient of the electrode material.

Rough guidelines for $Z_0$ and $n_{\text{RF}}$ are also useful. In lossless transmission lines, they are approximately

$$Z_0 = \sqrt{\mathcal{L}/\mathcal{C}},\ n_{\text{RF}} = c\sqrt{\mathcal{L}\mathcal{C}}, \tag{6}$$

where $\mathcal{C}$ and $\mathcal{L}$ are the line's capacitance and inductance per unit length.

## III. Lifting The Bandwidth-Voltage Trade-Off

The above analytical expressions serve as yardsticks against the more complex structures simulated numerically in this work. Equation (1) suggests three efficiency levers to lower $V_\pi \cdot L$, i.e., reducing $G$, raising $n_g$, and raising Γ. However, altering these levers can degrade a parameter associated with the EO bandwidth or increase the optical loss.

Many approaches have been pursued to alleviate this bandwidth-voltage trade-off and achieve high-bandwidth, low-bias MZMs. Some of the key approaches are summarized in the schematics of Figs. 1(b)-(f), while Fig. 1(a) depicts the simplest benchmark structure based on TFLN waveguides on oxidized silicon (Si) substrates and CPW traveling-wave electrodes.

Two seemingly straightforward refinements of the benchmark structure follow from scaling of $V_\pi$, i.e., narrowing the electrode gap and lengthening the device. Neither yields a net improvement on its own. A narrower gap brings the metallic layers into the edge of the optical modes and increases the optical propagation loss. It also negatively impacts the RF properties, as discussed later. A longer device likewise lowers $V_\pi$ for a fixed $V_\pi \cdot L$, but RF attenuation and velocity walk-off accumulate with electrode length (Eqs. (4) and (5)), so a penalty is paid in bandwidth; folding the optical path compacts the overall device footprint [10-15], but without lifting the basic constraint, since the RF penalties scale with the total unfolded length.

### *A. Electro-optic overlap*

Two approaches can raise Γ. In the first approach, the optical mode is moved toward the RF field. Since the modulating field is strongest near the electrode edges, displacing the waveguide off-center toward the signal electrode by an offset, $\Delta D$, as depicted in Fig. 1(b), places the mode in the high-field region and raises Γ. The penalty is increased optical absorption, as the mode tail now reaches the electrodes' metal. A thin low-$\varepsilon$ buffer between the electrode and the LN (e.g., silicon dioxide (SiO2) in Fig. 1(b)) suppresses the absorption. The drawback is a slight decrease in Γ and a slight increase in $V_\pi$, but overall, the performance improves.

The concept of non-centered placement of the optical waveguide with respect to the electrode can be traced back to bulk LN guided-wave devices [16], [17] and has been applied to the thin-film platform, where tighter confinement permits much smaller electrode-to-mode separations [18]. EO bandwidths approaching 400 GHz are theoretically attainable in millimeter-scale devices, but at the expense of high $V_\pi$ of ~ 13 V [19]. Experimental results on this scheme demonstrate bandwidths of 170 GHz at a $V_\pi$ of 6.6 V in 5-mm-long devices [20]. Introducing optical isolation trenches can also increase field confinement and reduce optical loss [21].

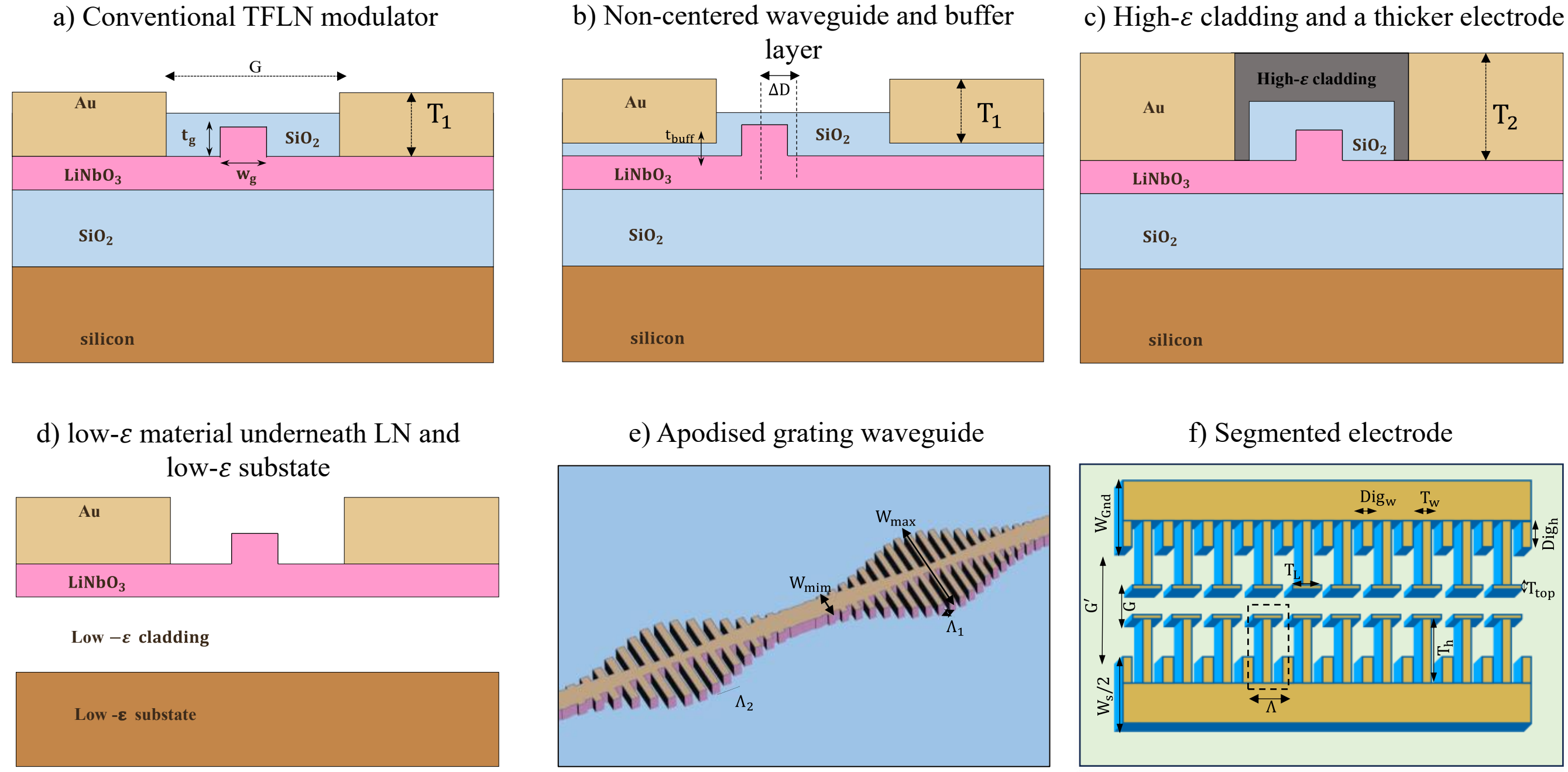

Fig. 1. Different mechanisms are designed and optimized together to enhance the electro-optic modulator.

In another approach to increase Γ, the top cladding layer of the TFLN waveguide is made of a material with a higher $\varepsilon$, compared to the standard $SiO_2$ choice (Fig. 1(c)). Example materials are aluminum oxide, hafnium dioxide, and glycerol, as elaborated later in Section V.A. In any case, the high-$\varepsilon$ cladding induces a higher vertical electric field across the dielectric boundary and into the core through the continuity of the normal displacement field [18], [22], [23]. This approach can be paired with the above non-centered waveguide approach, as well as thicker electrodes for further performance enhancement [18], [23]. Furthermore, replacing the metal electrode with a transparent conducting oxide can remove the gap-versus-absorption trade-off at the electrode surface [24]. Disadvantages of the high-$\varepsilon$ cladding approach are the processing challenges of unconventional materials and their potential incompatibility with foundry production.

### *B. Capacitively-loaded traveling-wave electrodes*

A narrower gap, $G$, reduces $V_\pi \cdot L$ but squeezes the signal current into a smaller cross-section and crowds it along the electrode faces, thereby raising the line resistance at high frequencies. It also pulls the RF field into the lithium niobate and the surrounding dielectric claddings, which increases $C$. Furthermore, concentrating the field in a tighter gap draws more of it into the high-$\varepsilon$ material, which slows the microwave wave and raises its effective index. The effect is mild over low-$\varepsilon$ surroundings but steep over a high-$\varepsilon$ substrate, where the microwave wave falls out of step with the optical wave and the bandwidth suffers. In addition, the extra capacitance of a tighter gap drops $Z_0$ below 50 Ω, and restoring the match requires reshaping the electrodes, which in turn feeds back on the loss and the velocity, so the three penalties cannot be tuned independently.

Abandoning the traditional CPWs for capacitively-loaded electrodes considerably alleviates the voltage-bandwidth tradeoff. Capacitive loading of the electrodes is implemented using periodically segmented electrode structures, e.g., the T-shaped teeth shown in Fig. 1(f). In ground-signal-ground (GSG) electrodes utilized for push-pull configuration, low RF loss (hence high bandwidth) requires a wide signal electrode. However, as this width increases in CPWs, the electrode gap, $G$, must also increase to maintain $Z_0 = 50\ \Omega$ [25]. As discussed, a larger $G$ reduces modulation efficiency and undermines the main motivation for thin-film EOMs.

Introducing T-shaped periodic structures between the ground and signal electrodes prevents electrical current crowding at the edge of the electrodes, hence less conductor ohmic loss contribution to the RF loss, and a small $G$ can be afforded. The challenge here is that $C$ increases with reduced $G$, hence $n_{\mathrm{RF}}$ (see Eq. (6)) deviates from $n_{\mathrm{eff}}$. The low-$\varepsilon$ substrate approach of Section III.C is one solution to retain velocity matching. Another approach is to enhance $n_{\mathrm{eff}}$ in slow-light structures, as discussed in Section III.D.

A complementary enhancement method is increasing the metal thickness, since the skin-effect conductor loss falls with electrode thickness until the thickness exceeds a few skin depths. Thick-plated metals, bimetallic stacks, dual-layer loading, and flip-chip geometries each exploit this scaling [26-29]. Together, these electrode strategies have delivered sub-2-V operation, in some cases with dual-output and high-extinction functionality, at bandwidths of roughly 100 to 160 GHz [30], [31].

An extra challenge with standard T-shaped segmented electrodes is that $\mathcal{C}$ increases, while $\mathcal{L}$ slightly decreases, thus $Z_0 = 50\ \Omega$ is difficult to achieve. Introducing an interdigitated structure can solve this trade-off, i.e., inserting conductive digits interleaved between the T-electrodes [8].

### *C. Low-ε substrates and suspended waveguides*

The microwave field penetrates the substrate to a depth comparable to $G$ and contributes to $\alpha_{\mathrm{RF}}$. Particularly, the RF field in Si substrates suffers from residual carrier (electrons and holes) loss that grows with frequency, while materials like quartz and silica have a much lower loss tangent [32]. However, employing such dielectric substrates disturbs the velocity matching, since the low-$\varepsilon$ surround gives the unloaded electrode a microwave index well below the optical group index. Therefore, low-$\varepsilon$ substrates work better in tandem with the segmented electrodes described in Section III.B. The two are complementary: the low-$\varepsilon$ substrate lowers the unloaded microwave index well below $n_{\mathrm{g}}$, and the capacitive loading then raises $n_{\mathrm{RF}}$ back toward it, recovering velocity matching at narrow gaps. TFLN EOMs on quartz substrates have been demonstrated [30], [33]. It can be argued that a disadvantage is the unconventional choice of substrate material for standard processing and potential heterogeneous integration with silicon photonics. Alternatively, partially air-clad TFLN waveguides on Si substrates have been demonstrated [34]. The drawback here is the more complicated processing steps to fabricate the suspended waveguides and their mechanical robustness.

### *D. Optical group-index enhancement*

The inverse dependence of $V_\pi$ on $n_{\mathrm{g}}$ in Eq. (1) points to slowing the optical wave to lower $V_\pi$. A slower optical field lingers longer in the modulating region and accumulates more phase shift per volt. Slow-wave structures built from Bragg gratings (e.g., see Fig. 1(e)) and photonic crystals raise $n_{\mathrm{g}}$ [8], [25],[35–38]. Coupled-Bragg-resonator and cascaded apodised grating waveguides reach group indices several times that of an ordinary rib waveguide over usable passbands [35], [38]. Single-grating slow-light modulators reaching a $V_\pi.L$ of 0.21 V·cm with bandwidths beyond 110 GHz have been reported [36]. However, the devices in this report operate at the band-edge transition, and thus have very limited optical bandwidth. Cascading a series of apodized Bragg gratings, separated by $\pi$-phase shifters, offers an optical passband of several nanometers [38]. In another approach, cascading two apodized grating waveguides with different Bragg wavelengths opens a smoother passband with up to 10 nm optical bandwidth [25].

Matching the slow optical wave, however, requires slowing the microwave wave as well. This can be readily achieved by incorporating the segmented electrodes discussed in Section III.B. Indeed, one advantage of slow-light modulators is that there will be no need for low-$\varepsilon$ substrates, i.e., simultaneous slow optical and RF waves can be demonstrated on standard TFLN on Si substrates. Co-designing cascaded slow-wave grating waveguides with velocity-matched interdigitated T-electrodes on standard Si substrates offers attainable $V_\pi$ around 1.25 V and EO bandwidths above 100 GHz [8]. Such electrodes with $n_{\mathrm{RF}}$ of up to 4.5 and $Z_0 \approx 50\,\Omega$ up to 67 GHz RF bandwidths have been measured [25].

From the microwave perspective, the electrode must be redesigned to achieve an equally high microwave index through capacitive loading. The T-rails add a capacitance per unit length $\Delta\mathcal{C}$ per arm, while the inductance stays essentially unchanged compared to the unloaded line; hence, the microwave index scales as $\sqrt{\mathcal{L}_0(\mathcal{C}_0 + 2\Delta\mathcal{C})}$ and the impedance as $\sqrt{(\mathcal{L}_0/(\mathcal{C}_0 + 2\Delta\mathcal{C}))}$ [39]. The same loading that slows the microwave to match the optical group index therefore also depresses the impedance below 50 Ω. Reducing the unloaded capacitance $\mathcal{C}_0$ with a low-$\varepsilon$ substrate and an air undercut relieves this conflict, so the fill factor can stay at 90% for low drive voltage while the interleaved digits restore the impedance. The period is set by a second constraint, because a periodically loaded line also behaves as a filter and must be short enough to keep the cut-off frequency well above the operating band [39]. Once velocity matching is preserved, the bandwidth is limited by conductor loss. A thicker electroplated gold reduces this loss. The resulting electrode of Fig. 1(f) recovers velocity and impedance matching at narrow gaps, as demonstrated on quartz substrates with a predicted 3-dB bandwidth exceeding 300 GHz for undercut variants [32].

Another challenge of slow-light waveguides is that grating waveguides are generally more lossy than standard $z$-invariant waveguides due to higher scattering through etched sidewalls as well as radiation [40]. Also, as the group index increases, the optical passband narrows to a couple of nanometers [35].

## IV. General Co-designed Structure

An augmented EOM design can be imagined, where the five enhancement routes introduced in Section III augment each other in the same device. Such a device is shown in Fig. 2. The slow-wave grating waveguide is taken from the benchmark design and held fixed throughout. Its rib width, etch depth, corrugation widths, super-Gaussian apodization, and pitch detuning are not re-optimized here, so the group index stays at $n_{\mathrm{g}} = 4.3$ for every structure reported below.

CST Design Studio was used for RF modeling, and Ansys Lumerical was used for optical modeling in this work. The slow waveguide, shown in Fig. 1(e), follows the cascaded dual-apodized-grating method of Haefner *et al.* [8], where two Bragg gratings of different pitch open a low-dispersion slow-light passband between their stopbands, and the group index is set by a single variable, the pitch detuning $\Delta\Lambda = |\Lambda_1 - \Lambda_2|$.

Specifically, the optical waveguide consists of a rib with $w_g$ = 950 nm width, and a $t_g$ = 220 nm etch depth on a 600-nm X-cut film, with grating corrugation widths of $w_{min}$ = 0.56 µm and $w_{max}$ = 1.02 µm and super-Gaussian apodization. A detuning of $\Delta\Lambda$ = 39 nm yields $n_g$ = 4.3 with a near-2 nm passband [8].

Although these dimensions are taken from the benchmark structure, which uses a Si substrate with $SiO_2$ bottom cladding and air top cladding, they remain applicable to our augmented structure because the high-$\varepsilon$ upper cladding and the air undercut shift the optical effective index in opposite directions and largely compensate each other. Replacing the bottom oxide with air lowers $n_{eff}$ by about 0.011, while the $SiO_2$ buffer and glycerol cladding above the waveguide raise it by a comparable amount, leaving a net shift below 0.001. Furthermore, in all studied devices, an identical 150-nm $SiO_2$ buffer layer beneath the electrodes suppresses the metal absorption that the $\Delta D$ displacement would otherwise incur. In the following, our optimization strategy to achieve the general co-designed structure in Fig. 2 is explained, and more specific details for the low- and high-$\varepsilon$ material choices are given. The impact of different choices of high-ε layers is discussed first.

## V. Results and Discussion

### A. Overlap integral and bias enhancement

Two independent levers act on the overlap factor, Γ. The waveguide can be shifted toward the signal electrode by a displacement $\Delta D$, as depicted in Fig. 1(b), and a high-$\varepsilon$ layer can be applied as a top cladding above the $SiO_2$ cladding surrounding the waveguide, as depicted in Fig.1(c).

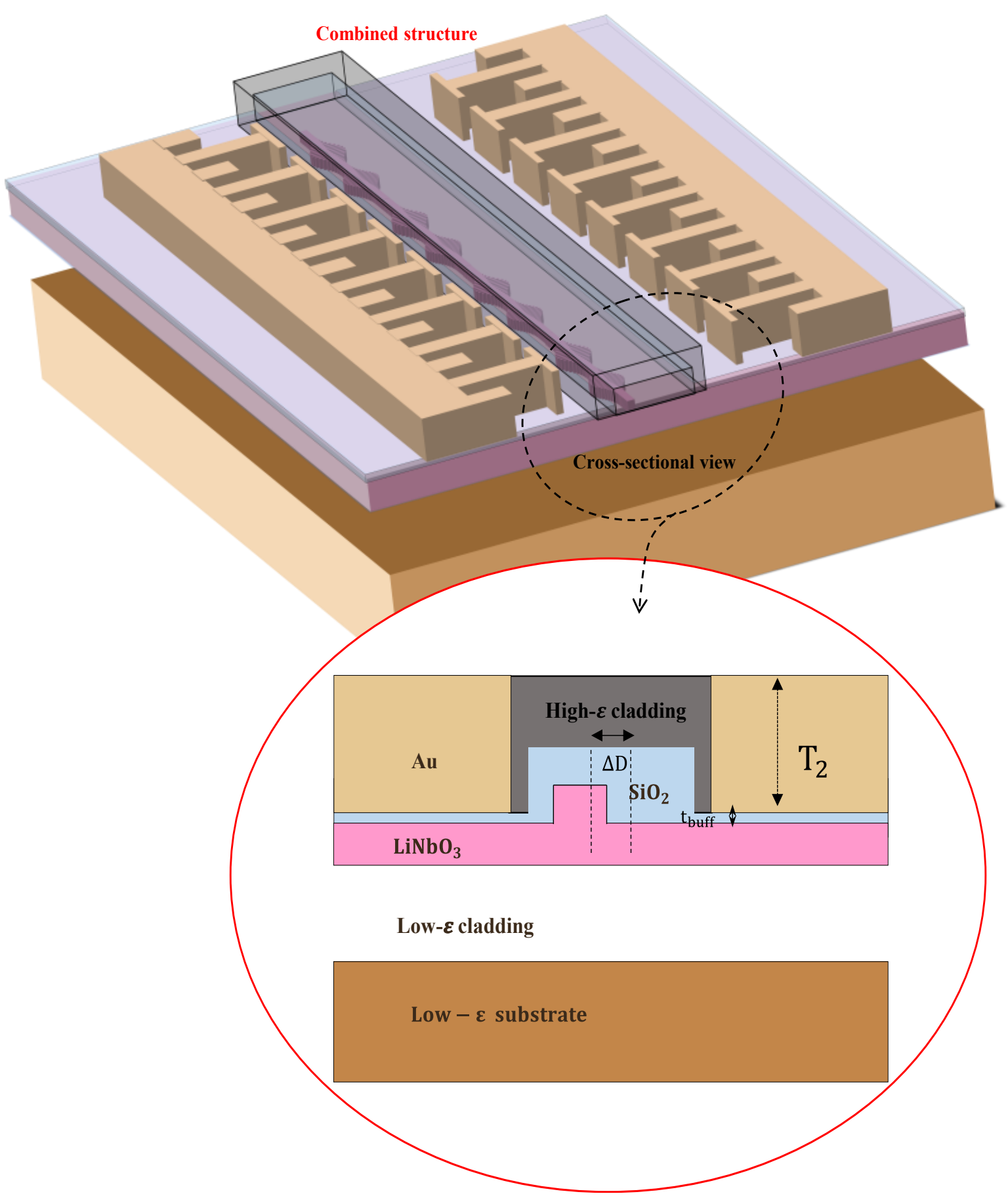

Fig. 2.Schematic of the co-design EOM structure

Figure 3 plots Γ versus the waveguide shift $\Delta D$ toward the signal electrode for different upper-cladding materials. Γ rises monotonically and linearly with $\Delta D$ in all cases. The plots are ordered by cladding $\varepsilon$, from air ($\varepsilon$ = 1) up to glycerol ($\varepsilon$ = 44). A higher $\varepsilon$ draws more of the electric field into the waveguide and lifts the whole trend. The best result is obtained with glycerol at $\Delta D = 0.8$ µm, at which $\Gamma = 0.74$, and is thus used in the following optimized co-designed structure.

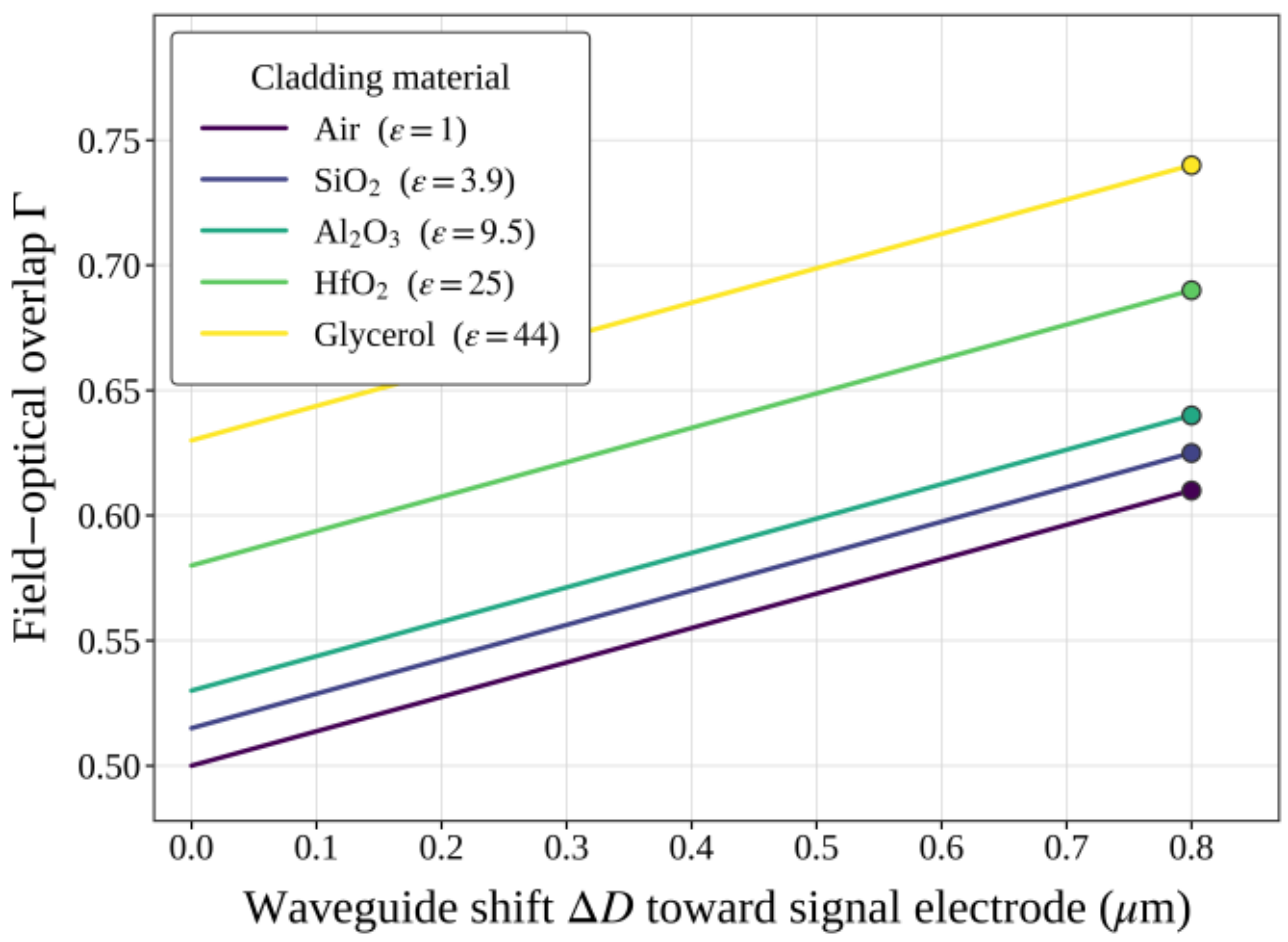

Fig. 4.Field-optical overlap Γ as a function of waveguide displacement *ΔD* toward the signal electrode for five upper-cladding materials, from air (ε = 1) to glycerol (ε = 44).

*B. Transmission line*

The segmented electrode and its geometrical dimensions are depicted in Fig. 1(f). This builds on the interdigitated T-electrode concept of Haefner *et al.* [8], [25], in which T-rails supply the capacitive loading, and interleaved digits restore the inductance, so the microwave index and impedance are tuned nearly independently. The concept is retained here but re-dimensioned, since the low-$\varepsilon$ substrates leave the unloaded line with far less capacitance than on Si, and the loading must supply a larger share of the index increment to hold the velocity mismatch near zero.

For example, the period, Λ, is 20 µm for a quartz substrate and air undercut, compared to 25 µm in Ref. [8]. An 18 µm active rail holds the duty cycle at 90% with a 2 µm inter-rail spacing. The signal electrode width, $W_s$, is widened from 105 µm in Ref. [8] to 120 µm, which deepens the loading and lowers the series resistance of the trunk, with a ground electrode width of $W_{gnd} = 100$ µm and the unloaded gap is set at $G' = 26$ µm, which is narrower than on Si substrate in Ref. [8] because the light dielectric environment already leaves the framework inductance-rich. These choices do not affect $V_\pi$ since the modulating field is confined to the loaded rail gap, $G$. The digit width $Dig_w$, the T-stem width, $T_{top}$ and the T-stem length, $T_L$ are fixed at 6, 2, and 18 µm, leaving the T-head width, $T_w$ and the digit length $Dig_h$ as the two closure levers, which converge near $T_w \approx 8.5$ µm and $Dig_h \approx 25$ µm for the present low-$\varepsilon$ cladding beneath the waveguide. The T-height, $T_h$, follows from the rail geometry once these are set. Finally, the electrode thickness, $T_2$, is 1.0 µm of electroplated gold, from $T_1 = 500$ nm to reduce the skin-effect conductor loss and add sidewall capacitance between the facing rails. The quartz substrate is 500 µm thick, and the air undercut depth is 4 µm, retaining low attenuation while preserving the dielectric beneath the rails that the high microwave index requires.

Figure 4 reports the index mismatch and the RF attenuation for a quartz/air/glycerol (solid) against the benchmark case of $Si/SiO_2$/air(dashed) for $G = 5.0$ µm. In panel (a), both lines are tuned onto $n_g$ of 4.3 at low frequency, but their high-frequency behavior separates: on quartz, $n_{RF}$ remains within 0.06 of the 4.30 target up to 219 GHz, whereas on Si it climbs to 4.47. On quartz, the rise is only the residual dispersion of the periodically loaded line, about 1.4 percent at 219 GHz for the 20-µm segmented electrode period. In the Si case, $n_{RF}$ rises much faster with frequency, since the RF mode migrates deeper into the high-$\varepsilon$ substrate. Effectively, for the same interaction length, the velocity-mismatch-limited 3-dB bandwidth is more than three times higher in quartz/air/glycerol compared to $Si/SiO_2$/air.

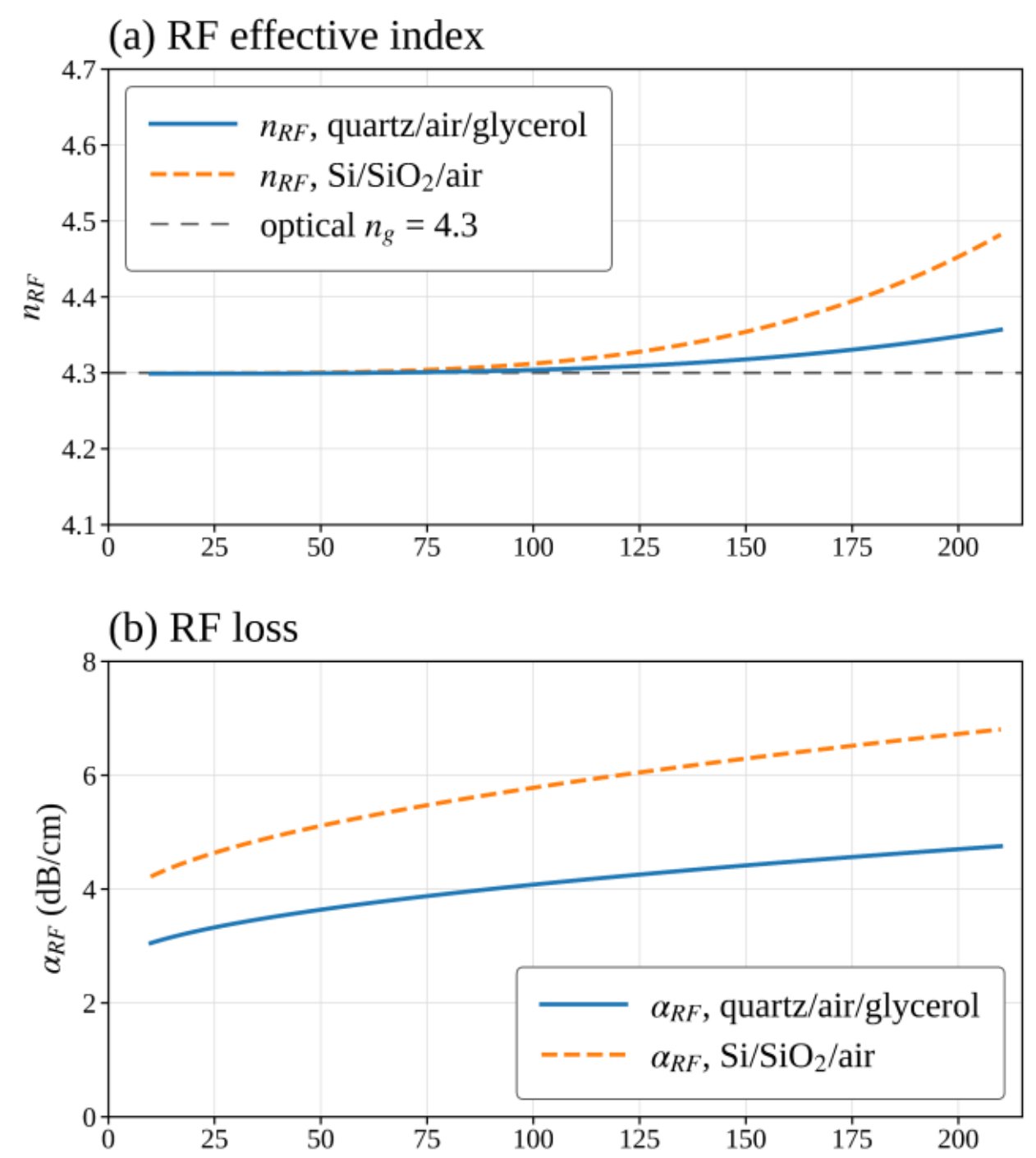

Fig. 3.RF-engineered metrics of the co-designed segmented transmission line versus frequency, for quartz/air/glycerol (solid) and Si/SiO2/air (dashed). Panel (a) is matched to the optical group index $n_g$ = 4.3, and panel (b) shows the RF loss $\alpha_{RF}$.

Figure 4(b) shows the corresponding attenuation, growing more slowly than the $\sqrt{f}$ dependence of CPWs. Specifically, the loss grows from 3.0 dB/cm at 10 GHz to 4.7 dB/cm at 219 GHz on quartz/air/glycerol, compared with 4.2 to 6.8 dB/cm on $Si/SiO_2$/air. Meanwhile, by tuning the inductance of the line, the 50-Ω impedance matching is maintained in both cases.

Figure 5 isolates the effect of the substrate on the RF loss and motivates the choice of gap, $G$. Panels (a) and (b) plot $\alpha_{RF}$

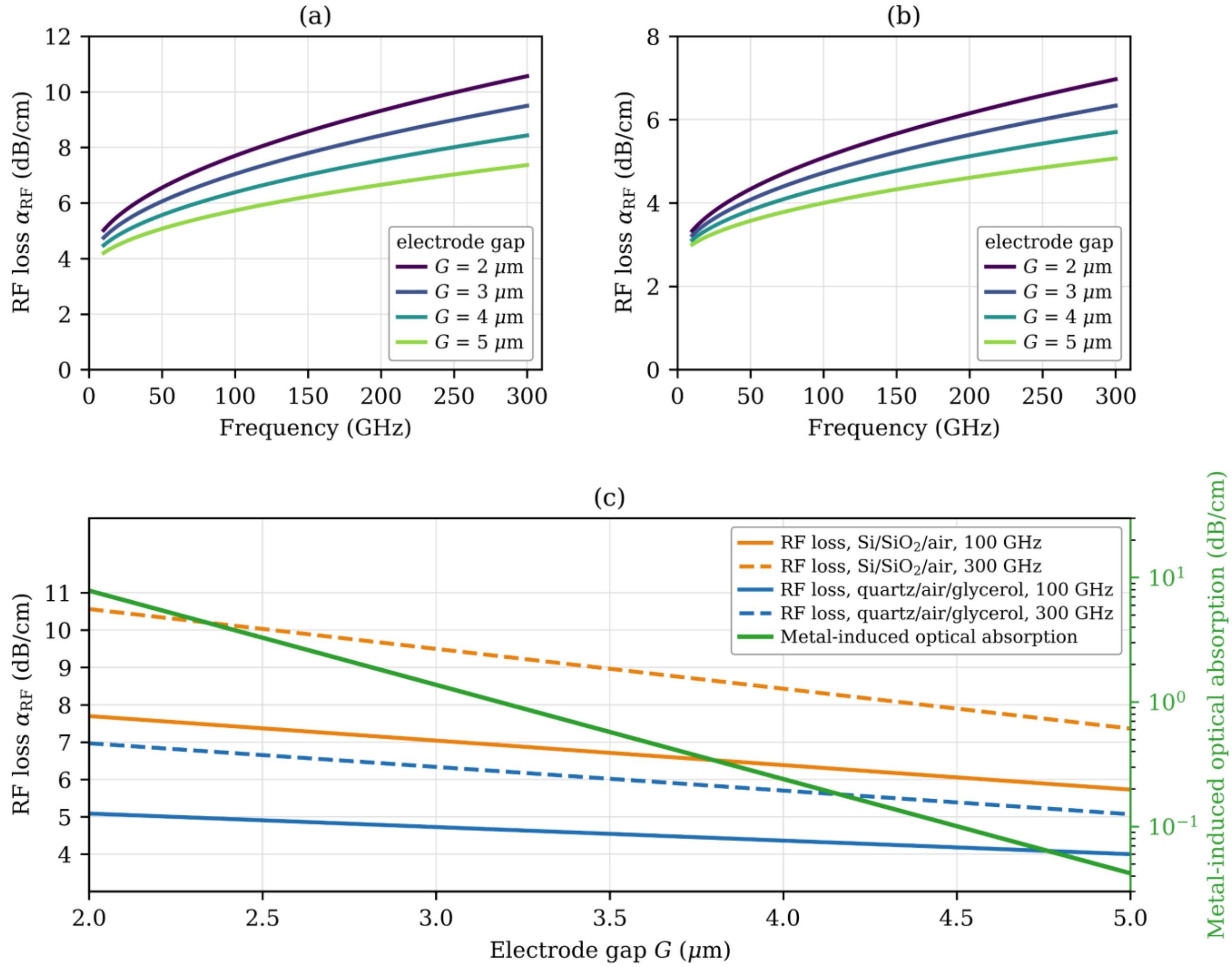

Fig. 5. $\alpha_{RF}$ versus frequency for electrode gaps of 2-5 μm on (a) the benchmark case of Si substrates and an $SiO_2$ bottom cladding layer and air top-cladding (Si/$SiO_2$/air); (b) quartz substrate with air undercut and glycerol upper cladding (quartz/air/glycerol); (c) RF loss (left axis) and metal-induced optical absorption (right axis) versus electrode gap.

versus frequency for electrode gaps of 2.0 to 5.0 μm on Si substrates and an $SiO_2$ bottom cladding layer and air top cladding (Si/$SiO_2$/air); and on a quartz substrate with air undercut and glycerol upper cladding (quartz/air/glycerol). At 300 GHz, the loss in the Si/$SiO_2$/air case reaches 7.2 to 10.6 dB/cm across the gap range, whereas for the quartz/air/glycerol case, the loss is 5.1 to 7.1 dB/cm. Panel (c) plots the RF loss (left axis) and the metal-induced optical absorption (right axis) against $G$. Both losses grow as the gap narrows. The 10-mm-long quartz/air/glycerol device may be an optimal case, since it delivers a 3-dB bandwidth of > 200 GHz and a $V_\pi < 1$ V (0.84 V). The highest bandwidth of 357 GHz corresponds to a $V_\pi$ of 3.36 V. For the Si/$SiO_2$/air case, an RF bandwidth of almost 100 GHz is attainable in 20-mm-long MZMs with $V_\pi = 0.63$V. Above 200 GHz operation is feasible in this case too, but with a high $V_\pi$ of 5V.

### C. Electro-optic bandwidth

Figure 6(a) shows the EO response, $S_{21}$, as a function of frequency for different device lengths and for the quartz/air/glycerol (solid) and Si/$SiO_2$/air (dashed) cases, both with $G = 5.0$ μm. The legend gives the respective EO bandwidths and lengths. The corresponding DC $V_\pi \cdot L$ are 0.84 and 1.25 V·cm. As a better guideline, the bandwidths are plotted versus $V_\pi$ in Fig. 6(b). It is noted that this is the DC $V_\pi$, while the RF $V_\pi$ scales with the inverse of $|H(\omega_{\mathrm{RF}})|$.

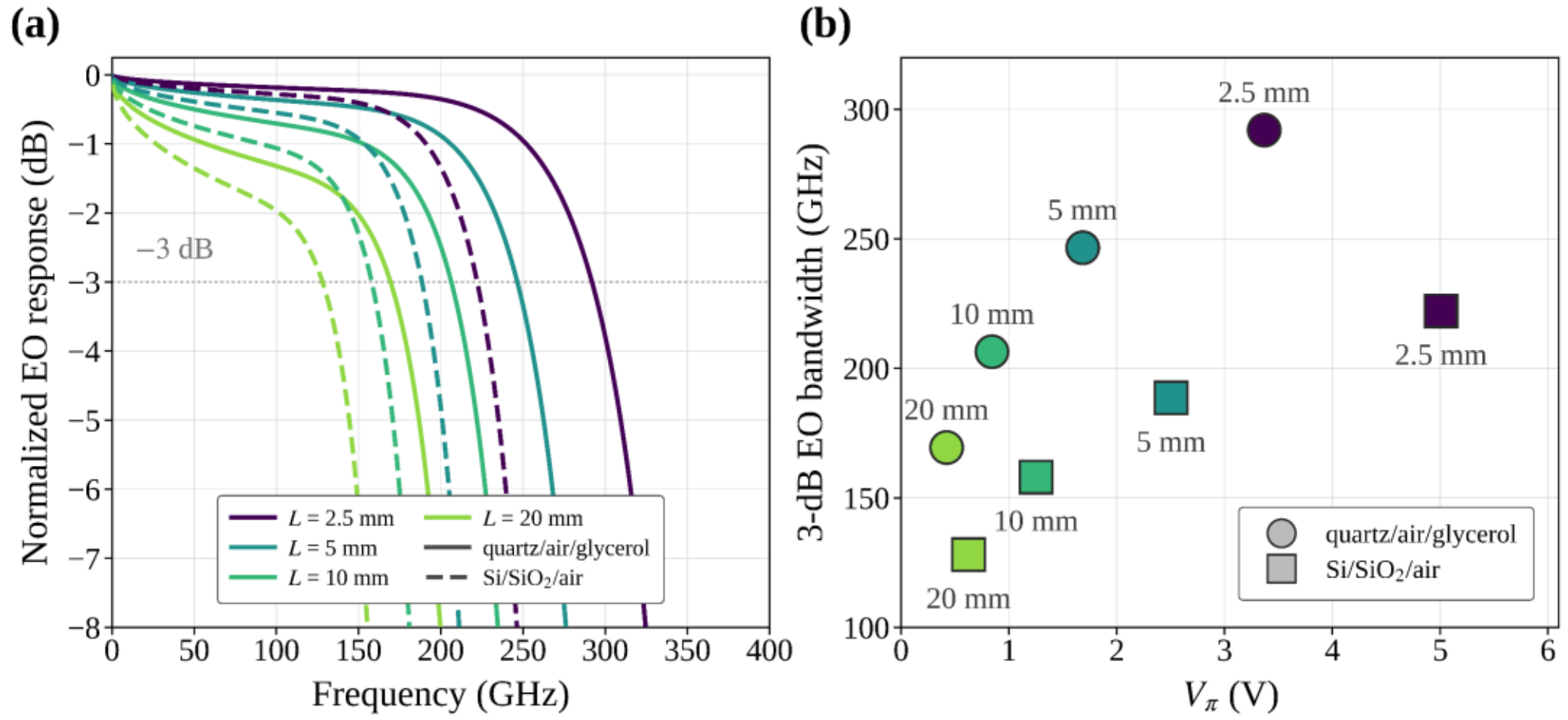

Fig. 6.(a) EO response, $S_{21}$, versus frequency for different modulator lengths in two cases of quartz/air/glycerol (solid) and $Si/SiO_2$/air (dashed). The corresponding 3-dB bandwidths are plotted versus the DC $V_\pi$ in (b).

## VI. Final Remarks on the Co-design and Comparison with the State of the Art

Figure 7(a) summarizes the additive evolution of the performance improvement in $V_\pi$ and EO bandwidth across six cases. Case 1 uses straight optical waveguides with CPW and thin ($T_1 = 500$ nm) electrodes (Fig. 1(a)). Case 2 adds a slow-wave grating waveguide (Fig. 1(e)) with CPW thin electrodes. Case 3 combines the slow-wave grating waveguide with a slow-wave thin electrode structure (Fig. 1(f)). Case 4 introduces an air bottom cladding on a quartz substrate (Fig. 1(d)). Case 5 shifts the waveguide toward the signal electrode and adds a buffer layer beneath the electrode (Fig. 1(b)). Case 6 adds an upper high-$\varepsilon$ cladding (glycerol) with thicker ($T_2 = 1$ μm) electrodes (Fig. 1(c)). This last case provides a pathway to an RF bandwidth above 200 GHz and $V_\pi$ < 1 V, as noted in the previous section. Evidently, combining displaced waveguides, a slow-wave optical grating, high-ε upper claddings, low-ε substrates, electrode thickening, and segmented loading can meet these metrics. Figure 7(b) compares the present design with reported measured (filled markers) and simulated (open markers) TFLN MZMs.

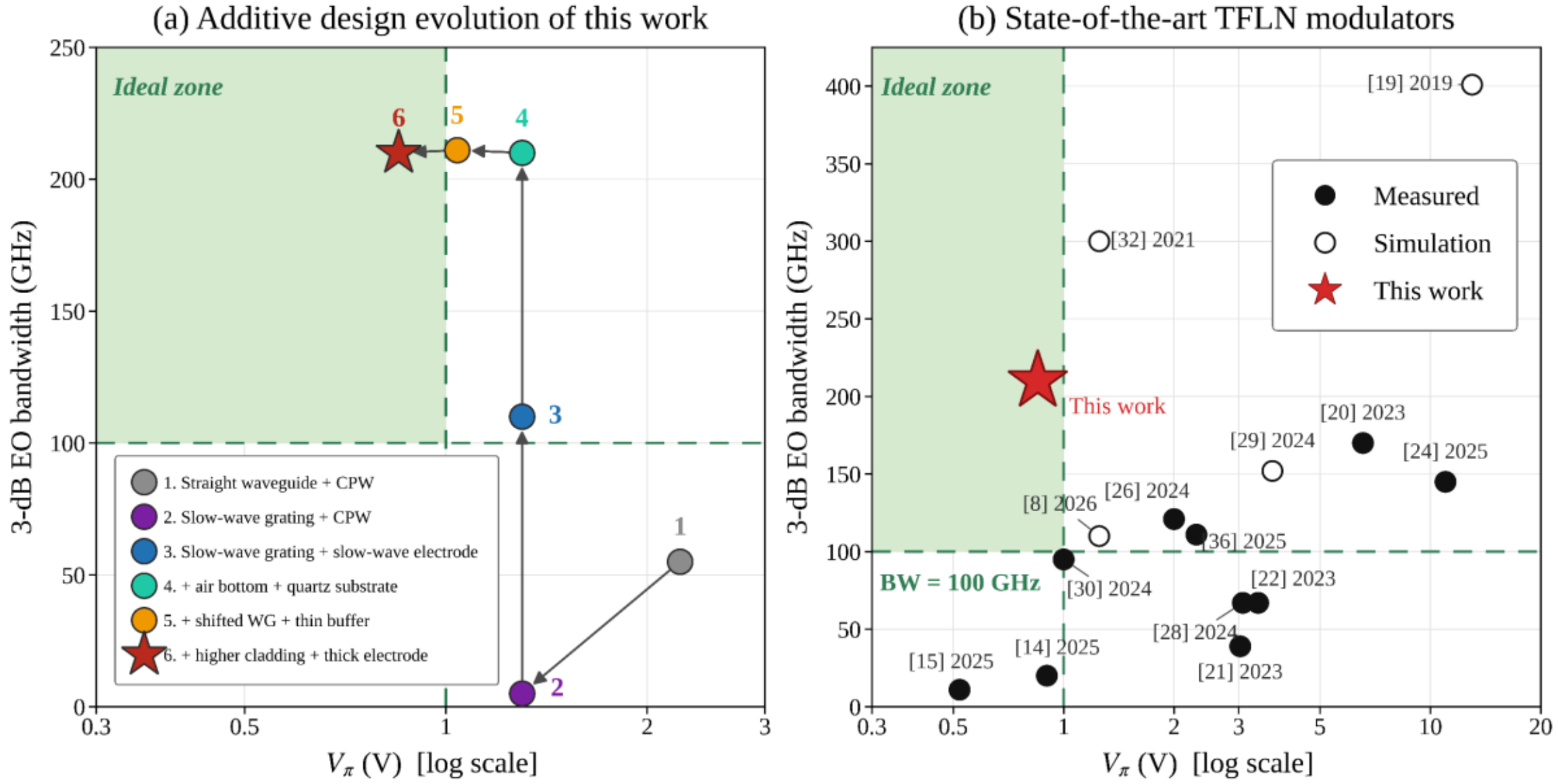

Figure 7.(a) Additive evolution: Comparison of the simulated 3-dB EO bandwidth and half-wave voltage across the six design cases, from straight waveguides with CPW electrodes to the fully optimized structure with a high-$\varepsilon$ upper cladding and thickened, segmented electrodes. (b) Comparison of the state of the art for reported TFLN MZMs (measured and simulated), and the present work. The shaded region marks a sub-1-V drive with a bandwidth above 100 GHz.

The reported devices follow a clear trade-off between drive voltage and bandwidth. Sub-volt operation has been demonstrated, but only at modest EO bandwidths, with approximately 10 GHz at 0.52 V [15] and 20 GHz at 0.9 V [14]. Conversely, the devices that exceed 100 GHz do so at several volts, such as 120 GHz at 2.0 V [26], 110 GHz at 2.1 V [36], 145 GHz at 11 V [24], and 170 GHz at 6.6 V [20]. Simulated designs push the bandwidth higher without resolving the trade-off, reaching 400 GHz at 13 V [19], 300 GHz at 1.2 V [32], and 110 GHz at 1.25 V [8].

The shaded region in Fig. 7(b), marking a sub-1-V drive with a bandwidth above 100 GHz, is therefore unoccupied. The present Case 6 co-designed structure, with 219 GHz and 0.84 V performance, lies well inside the shaded region and clears the bandwidth threshold by more than a factor of two.

This work focuses on modeling to explore the ultimate performance of TFLN MZMs, while the challenges of fabricating the complex quartz/air bottom cladding and encapsulating liquid glycerol at room temperature are beyond the scope of this work. The more standard Si/$SiO_2$/air case study eliminates these fabrication challenges and is more compatible with foundry production. As mentioned in Section V.C, operation for applications up to ~ 100 and 210 GHz bandwidths is feasible in this case, but with a higher $V_\pi$ of 0.63 and 5.0 V, respectively.

TABLE 1

Different design parameters and metrics for the quartz/air/glycerol case and $L = 10$ mm.

| Parameter | Design #1 | Design #2 | Design #3 |
|---|---|---|---|
| Optical group index, $n_g$ | 4.3 | 3.7 | 4.3 |
| Electrode gap, $G$ (μm) | 5.0 | 5.0 | 4.0 |
| Pitch detuning, $\Delta\Lambda$ (nm) | 39 | 43 | 39 |
| Rail-head width, $T_w$ (μm) | 8.5 | 6.7 | 6.7 |
| Digit height, $Dig_h$ (μm) | 25 | 21 | 24 |
| Half-wave voltage, $V_\pi$ (V) | 0.84 | 0.91 | 0.64 |
| 3-dB EO bandwidth (GHz) | 219 | 219 | 187 |
| Metal-induced optical loss (dB/cm) | 0.042 | 0.042 | 0.240 |
| Optical bandwidth (nm) | 3.3 | 5.1 | 3.3 |

Some further tweaks on Case 6 in Fig. 7(a) are discussed here. Table 1 compares three related designs, with the same 10-mm MZM arm length, in terms of the operating point, isolating the two principal voltage levers. Comparing designs #1 and #2, both at $G = 5.0$ μm, raising the group index from 3.7 to 4.3 through a smaller grating detuning lowers the half-wave voltage from 0.91 to 0.84 V, close to the inverse group-index scaling of Eq. (1), while the bandwidth remains at 219 GHz; the cost appears instead in the optical passband, which narrows from 5.1 to 3.3 nm through the accumulated group delay.

In design #3, $G$ is reduced to 4.0 μm to obtain a lower half-wave voltage of 0.64 V. However, the metal-induced optical absorption rises nearly six-fold, from 0.042 to 0.240 dB/cm, and the bandwidth falls to 187 GHz through the higher conductor loss and microwave index of the narrower line. In each case, the electrode is re-converged onto its target: the lighter loading of design #2 is closed with a narrower rail head and shorter digits ($T_w \approx 6.7$ μm, $Dig_h \approx 21$ μm), while the added rail capacitance of the 4 μm gap in design #3 is compensated by retreating the rail head at a nearly unchanged digit length. The group index, therefore, trades the drive voltage against the optical passband at essentially fixed bandwidth, whereas the gap trades the drive voltage against optical absorption and a modest bandwidth penalty. The operating point of design #1 adopts the largest group index and gap for which both costs remain within the targeted voltage and bandwidth values.

## VII. CONCLUSION

In a thin-film lithium niobate traveling-wave Mach-Zehnder modulator, the voltage-length product and the bandwidth are governed by the same physical parameters, so three major voltage-reduction levers, namely electrode gap reduction, group-index enhancement, and RF-optical overlap enhancement, each lower the voltage-length product but at the cost of a bandwidth penalty. Different approaches have been pursued to alleviate this tradeoff, some of which are: (a) Shifting the position of the optical waveguide off-center with respect to the ground and signal electrodes; (b) High-$\varepsilon$ upper cladding material with thick electrodes; (c) Low-$\varepsilon$ substrate material; (d) Slow-light waveguides; (e) Segmented RF electrodes. This work presents a comparative study of these approaches, how each enhances the performance of the concerned modulators, and what the ultimate performance would be of a co-designed optimized structure that exploits all of these approaches simultaneously. This joint design comprises quartz substrates with air undercut waveguides and a glycerol upper cladding layer. The optimized design lowers the unloaded line index, reduces the conductor loss, and provides the loading headroom for the electrode to track the slow-wave grating at a group index of 4.3, while the 1-μm electroplated gold at the 5-μm gap and the glycerol cladding raise the RF-optical overlap to 0.74. The resulting 10-mm device with a 0.8 μm waveguide displacement attains a half-wave voltage of 0.84 V, a 3-dB EO bandwidth of ~ 220 GHz, and an optical bandwidth of 3.3 nm with negligible metal-induced loss of ~ 4 dB/m. This work focuses on the ultimate performance limit of thin-film lithium niobate modulators. While the studied modulators can in principle be fabricated with standard cleanroom processing techniques, practical challenges and considerations – such as air-undercut quartz substrates and encapsulating glycerol, which is in liquid form at room temperature – are beyond the scope of this work.

ACKNOWLEDGMENT

This work was supported by the National Science Foundation Industry-University Cooperative Research Centers IUCRC EPICA Program.